\documentclass[aps,pre,10pt,twocolumn,superscriptaddress]{revtex4-2}
\usepackage{graphicx}
\usepackage{float}
\usepackage{epstopdf}
\usepackage{verbatim}
\usepackage{xcolor}
\usepackage{amsmath}
\usepackage{amssymb}
\usepackage[utf8]{inputenc}
\usepackage{graphicx,float,hyperref}
\usepackage{natbib}

\usepackage{calligra}  % For calligraphic font
\usepackage{graphicx,latexsym} 

\begin{document}
	
	\title{{\large {\Large Optimal Performance of an Asymmetric Quantum Harmonic Otto Engine and Refrigerator}}}{\normalsize {\large }}
	
	\author{Monika}
	\email{monika.ph.23@nitj.ac.in}
	\affiliation{Department of Physics, Dr B R Ambedkar National Institute of Technology Jalandhar, Punjab-144008, India}
	
	\author{Shishram Rebari}
	\email{rebaris@nitj.ac.in}
	
	\affiliation{Department of Physics, Dr B R Ambedkar National Institute of Technology Jalandhar, Punjab-144008, India}

\begin{abstract}
	We study a quantum Otto cycle operating with a time-dependent harmonic oscillator as the working material. We examine the asymmetry present between the two adiabatic processes of the Otto cycle, focusing on cases of sudden expansion and sudden compression. We analytically derive the efficiency and coefficient of performance for an asymmetric Otto cycle, employing the Omega function, which represents the balance between the maximum useful energy and minimum lost energy. Notably, our findings reveal that the efficiency (coefficient of performance) of an asymmetric engine (refrigerator) is higher during the sudden compression case compared to the sudden expansion case. Furthermore, we derive the results for the maximum work efficiency and observe that efficiency at the maximum Omega function consistently exceeds the maximum work efficiency. Finally, we compute the fractional loss of work in both cases to thoroughly examine the performance of asymmetric Otto engine.
\end{abstract}

\keywords{Quantum Otto cycle; harmonic oscillator; Omega function.}

	\maketitle
%\tableofcontents

\section{Introduction}

In the modern era, thermal devices have primarily contributed to technological and industrial development. Heat engines and refrigerators are prominent examples of thermal devices \cite{92370057,borgnakke2020fundamentals}. Heat engines utilize heat to generate mechanical work; in contrast, the refrigerator consumes work to extract heat from a cold reservoir. The Carnot efficiency of a heat engine is formulated as $\eta_{c}$ = 1 - $T_{c}/T_{h}$ while $\zeta_{c} = T_{c}/(T_{h} - T_{c})$ is the Carnot coefficient of performance (COP) where $T_{c}$ and $T_{h}$ are the temperatures of the cold and hot reservoir, respectively \cite{singh2018low,fermi2012thermodynamics}. Thermal devices can only achieve these Carnot bounds when they run infinitely slowly; therefore, the Carnot cycle has limited practical importance.

However, in actual practice, the performance of thermal devices is often lower than the theoretical optimum because of entropy production in irreversible processes, which are finite in time and have finite size constraints \cite{rolandi2023collective,dann2020quantum}. Therefore, transitioning to finite-time thermodynamics becomes essential that helps to determine the optimal performance of thermal devices \cite{kaur2024optimization,singh2020optimal,izumida2013coefficient}. Quantum thermal devices that utilize quantum resources can potentially exceed the Carnot bound \cite{klaers2017squeezed,manzano2016entropy}.
Thus, exploring quantum heat engines and refrigerators presents an exciting field of research, offering possibilities beyond the conventional scope of macroscopic thermodynamics \cite{shaghaghi2022micromasers,singh2023thermodynamic,rossnagel2014nanoscale,bera2024steady,shaghaghi2023lossy,singh2020multi,singhoptimal}.

Recently, the quantum Otto cycle has attracted considerable research attention \cite{thomas2011coupled,insinga2020quantum,saryal2021bounds,deffner2018efficiency,alecce2015quantum,nautiyal2024finite,nautiyal2024out}. The quantum Otto cycle offers a simple yet powerful framework for analyzing the performance of thermal devices.
Optimizing these devices is essential for making them more efficient and sustainable \cite{PhysRevLett.109.203006,singh2018feynman, PhysRevE.106.024137,abah2016optimal}. Energy conversion is a commonly used method to optimize thermal devices. We have a well-known optimization function for energy analysis, known as the Omega ($\Omega$) function, which exhibits the best agreement between the maximum useful energy and minimum lost energy for engines, while for refrigerators, it is a trade-off between the maximum cooling load and minimum lost load \cite{hernandez2001unified,zhang2020unified,kaur2021unified}. Unlike the ecological function, which incorporates entropy through the environment's temperature, the Omega function excludes entropy and is independent of the environmental factors \cite{kaur2023optimal,liu2010ecological,fernandez2022optimization,singh2019three}.

Our work uses a quantum Otto cycle driven by a time-dependent harmonic oscillator as the working material \cite{kosloff2017quantum,kumar2023introduction,kaur2024performance}, and there is an asymmetry between the two adiabatic processes of the cycle. Based on the asymmetry introduced, we will discuss two key scenarios: sudden compression and sudden expansion. In the first scenario, the compression stroke is governed by a sudden-switch protocol, while the expansion stroke is adiabatic while the second scenario adopts the inverse approach \cite{Shastri_2022,singh2024asymmetric}. Firstly, we will find analytical expressions for efficiency at the optimal value of $\Omega$ function in the high-temperature regime for both scenarios. Then, we will discuss the fractional loss of work for both sudden compression and sudden expansion cases. Further, we will perform a similar task to determine the COP of the asymmetric Otto refrigerator. We will also do a comparative study of our results with those of the symmetric Otto engine and refrigerator that operates with both work strokes as either adiabatic or sudden-switch.

The organization of this paper is as follows:
Sec. \ref{sa} introduces the quantum Otto cycle with a time-dependent harmonic oscillator as the working material. In Sec. \ref{sb}, we derive analytical expressions of efficiency at the maximum $\Omega$ function. Sec. \ref{sb1} investigates the engine's performance during a sudden compression stroke, while in Sec. \ref{sb2}, we present our results for the sudden expansion stroke. Sec. \ref{sc} explores the concept of fractional loss of work. Further, we study asymmetric quantum refrigerator for sudden expansion and compression cases in Sec. \ref{sd}. We summarize and conclude our results in Sec. \ref{se}. 

\section{Quantum Otto Cycle} \label{sa}

\begin{figure}[t]
	\centering
	\includegraphics[width=8.2cm,height=5cm]{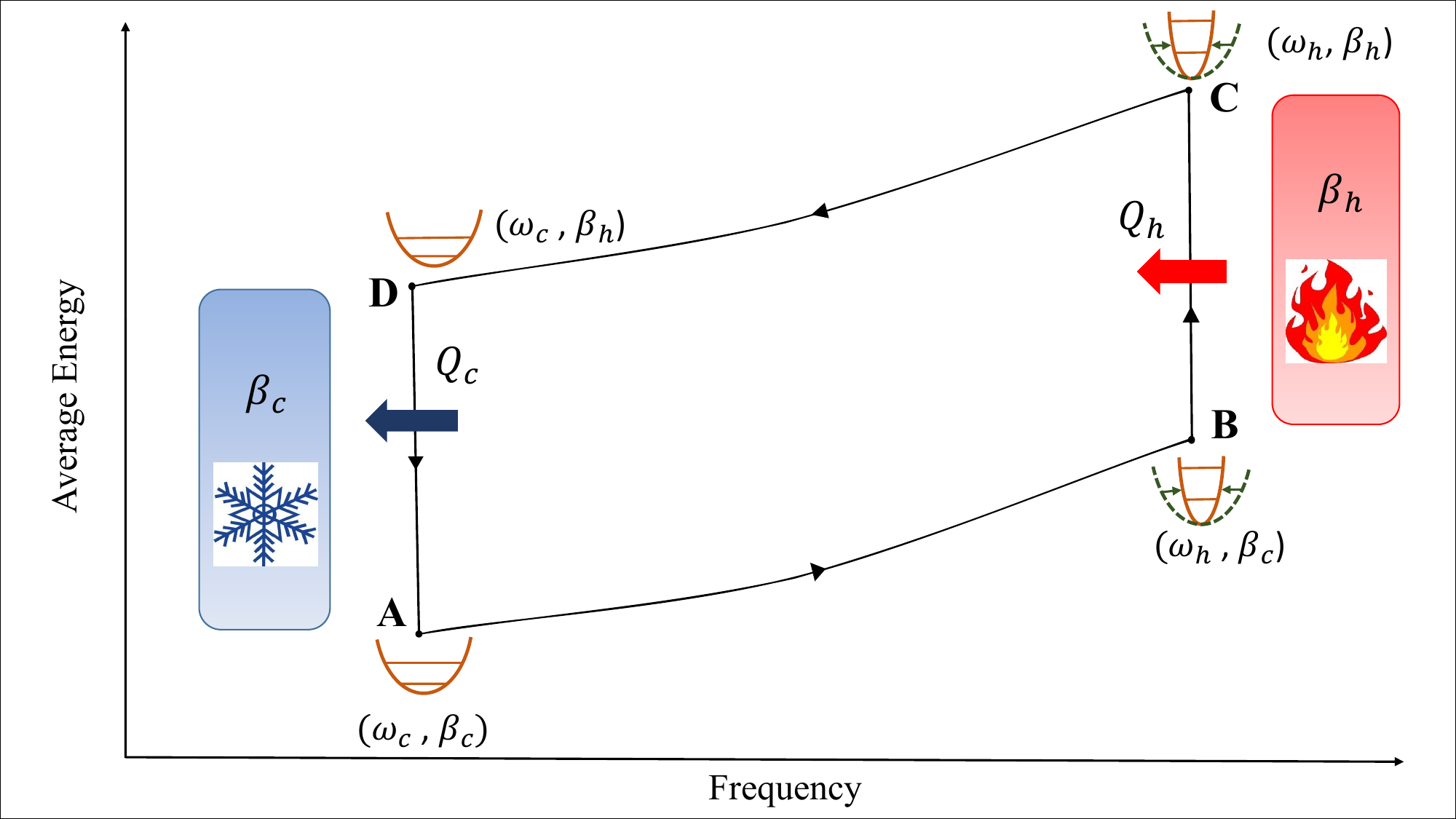}
	\caption{Representation of quantum Otto cycle employing a time-dependent harmonic oscillator as a working substance.}
	\label{fig:1}
\end{figure}

The quantum Otto cycle based on a time-dependent harmonic oscillator involves four sequential processes, which are briefly discussed as follows (see Fig. \ref{fig:1}) \cite{PhysRevE.106.024137,singh2020performance}: 

1). Adiabatic Compression (A $\to$ B): Initially, the system is at inverse temperature $\beta_{c}$. Then, with the help of an external agency, the working material undergoes adiabatic compression, and the oscillator's frequency changes from $\omega_{c}$ to $\omega_{h}$. Work is done on the system during this process.

2). Hot Isochore (B $\to$ C): Now, couple the system with a hot thermal reservoir having temperature $\beta_{h}$ = $(k_B T_{h})^{-1}$. The frequency $\omega_{h}$ of the harmonic oscillator remains fixed, with no work being done, only heat exchanges between the system and the reservoir.

3). Adiabatic Expansion (C $\to$ D): During this process, decouple the system from the thermal reservoir. The system expands adiabatically, and the oscillator's frequency gets reduced to its initial value $\omega_{c}$. In this stage, the system does the work.

4). Cold Isochore (D $\to$ A): In the final process, place the system in thermal contact with the cold reservoir having inverse temperature $\beta_{c}$ = $(k_B T_{c})^{-1}$. The heat exchanges between the system and reservoir, and the system returns to its initial thermal state A.

During these four stages of the cycle, the average energies $\langle H \rangle_{i}$ ($i = A, B, C, D$) of the oscillator are as follows ($\ k_{B}$ = $\hbar = 1$) \cite{PhysRevE.106.024137,singh2024asymmetric}:
\begin{align}
		\langle H \rangle_{A} &= \frac{\omega_{c}}{2} \coth \left(\frac{\beta_{c} \omega_{c}}{2}\right),\label{eq:1} \\
		\langle H \rangle_{B} &= \frac{\omega_{h}}{2}\lambda_{AB} \coth \left(\frac{\beta_{c} \omega_{c}}{2}\right), \label{eq:2}\\
		\langle H \rangle_{C} &= \frac{\omega_{h}}{2} \coth \left(\frac{\beta_{h} \omega_{h}}{2}\right), \label{eq:3}\\
		\langle H \rangle_{D} &= \frac{\omega_{c}}{2}\lambda_{CD} \coth \left(\frac{\beta_{h} \omega_{h}}{2}\right), \label{eq:4}
	\end{align}
where $\lambda_{i}$ ($i = AB, CD$) is the adiabaticity parameter and in general, $\lambda_i \geq 1$. For adiabatic cases, $\lambda_i$ = 1 and for sudden-switch cases, $\lambda_i$ = ($\omega^2_{c}+\omega^2_{h})/2 \omega_{c} \omega_{h} > 1$. 

The amount of heat transferred during hot and cold isochores are, respectively given by
	\begin{align}
		\langle Q \rangle_{h} &= \langle H \rangle_{C} - \langle H \rangle_{B} \nonumber \\
		&= \frac{\omega_{h}}{2} \left[\coth \left(\frac{\beta_{h} \omega_{h}}{2}\right) - \lambda_{AB} \coth \left(\frac{\beta_{c} \omega_{c}}{2}\right)\right], \label{eq:5} 
	\end{align}
\begin{align}
		\langle Q \rangle_{c} &= \langle H \rangle_{A} - \langle H \rangle_{D} \nonumber \\
		&= \frac{\omega_{c}}{2} \left[\coth \left(\frac{\beta_{c} \omega_{c}}{2}\right) - \lambda_{CD} \coth \left(\frac{\beta_{h} \omega_{h}}{2}\right)\right]. \label{eq:6}
	\end{align}

We follow a sign convention where we take input (output) work and heat to be positive (negative). Therefore, from the first law of thermodynamics, the net extracted work from a heat engine is given by $W_{ext} = Q_{c} + Q_{h}$.

\section{Asymmetric Quantum Otto Engine} \label{sb}

The efficiency of an engine is the ratio of extracted work and input heat and is given by 
\begin{equation}
	\eta = \dfrac{W_{ext}}{Q_{h}}.
\end{equation}
In this section, we study the optimization of Otto engine under the $\Omega$ function, which is balance between the maximum useful energy and minimum lost energy and is expressed as \cite{PhysRevE.106.024137}:  
\begin{equation}
	\Omega = 2W_{ext}-\eta_{max}Q_{h},\label{eq:7}
\end{equation}
where $\eta_{max}$ is the maximum attainable efficiency of an engine and is always either equal to or less than the Carnot efficiency, i.e., $\eta_{max}$ $\leq$ $\eta_{c}$. We consider asymmetry between the two adiabatic processes of the Otto cycle and will discuss two cases: sudden expansion and sudden compression \cite{monika2024asymmetric}. Here, we will study asymmetric quantum Otto engines only in high-temperature regimes as at low temperatures, the asymmetric Otto cycle does not function as a heat engine because the positive work condition fails to hold as discussed in Ref. \cite{singh2024asymmetric}. We initiate our study with the sudden compression stroke and then extend our investigation to the sudden expansion case.

\subsection{Sudden compression stroke} \label{sb1}

In this case, we consider the compression stroke (A to B) to be sudden while the expansion stroke (C to D) is still adiabatic and the adiabaticity parameters are thus $\lambda_{AB}$ = $(\omega^2_{c}+\omega^2_{h})/2 \omega_{c} \omega_{h}$ and $\lambda_{CD}$ = 1. In order to obtain the analytic results, we will discuss our model under the high-temperature limit. We obtain the amount of heat absorbed and the work extracted from the engine by setting $\coth \left(\beta_{i}\omega_{i}/2\right) \approx  2/\beta_{i}\omega_{i} $ ($i = c, h$):
\begin{equation}
	Q_{h} = \dfrac{1}{\beta_{h}}\left[1-\dfrac{\tau}{2}\left(1+\dfrac{1}{z^{2}}\right)\right]\label{eq:31},
\end{equation}
\begin{equation}
	W_{SC}^{HT} = \dfrac{1}{\beta_{h}}(1-z)\left(1-\dfrac{(1+z)\tau}{2z^{2}}\right),\label{eq:21}
\end{equation}
where $z =\omega_{c}/\omega_{h}$ is the compression ratio and $\tau$ = $\beta_{h}$/$\beta_{c}$. The corresponding efficiency expression for this case is as follows \cite{singh2024asymmetric}:
\begin{equation}
	\eta^{HT}_{SC} = \dfrac{(2z^{2}-\tau z-\tau)(1-z)}{z^{2}(2-\tau)-\tau}.\label{eq:8}
\end{equation}
We obtain the maximum attainable efficiency by optimizing Eq. (\ref{eq:8}) with respect to $z$, i.e., by setting $\partial$$\eta_{SC}^{HT}$/$\partial z = 0$ and comes up with the following cubic equation
\begin{equation}
	z^3 (2 - \tau) - 3 \tau z + 2 \tau^2 = 0. \label{a}
\end{equation}
We can not solve the above equation in terms of real radicals. Therefore, we solve this cubic equation by using the concept of casus irreducibilis and obtain the solution in terms of inverse trigonometric functions as (see Appendix \ref{appendixA}) \cite{singh2024asymmetric}:
\begin{equation}
	z^* = 2\sqrt{\dfrac{\tau}{2-\tau}} \cos\left[\dfrac{1}{3}\cos^{-1}\left(-\sqrt{\tau(2-\tau)}\right)\right].\label{b}
\end{equation}
\begin{figure}[t]
	\centering
	\includegraphics[width=8.6cm]{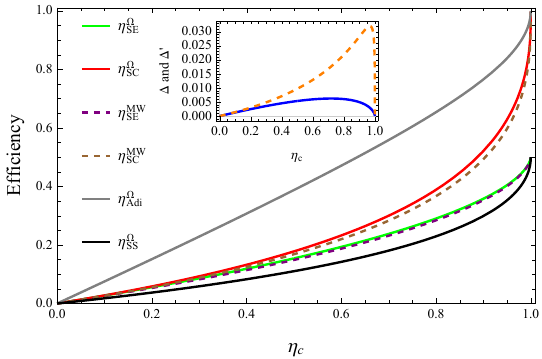}
	\caption{Plot of optimal efficiencies as a function of Carnot efficiency. Solid red and green curves represent $\eta_{SC}^{\Omega}$ (Eq. (\ref{eq:11})) and $\eta_{SE}^{\Omega}$ (Eq. (\ref{eq:16})) respectively while dashed brown and violet corresponds to $\eta_{SC}^{MW}$ (Eq. (\ref{32})) and $\eta_{SE}^{MW}$ (Eq. (\ref{33})) respectively. Solid gray and black curves represent efficiency at maximum $\Omega$ function for adiabatic ($\eta_{Adi}^{\Omega}$) and sudden-switch cases ($\eta_{SS}^{\Omega}$), respectively. In the inset, dashed orange curve represents $\Delta$ = $\eta_{SC}^{\Omega}$ - $\eta_{SC}^{MW}$ and solid blue curve represents $\Delta^{'}$ = $\eta_{SE}^{\Omega}$ - $\eta_{SE}^{MW}$.}
	\label{fig:2}
\end{figure}
The resulting expression of maximum attainable efficiency ($\eta_{SC}^{max}$) by using Eq. (\ref{b}) in Eq. (\ref{eq:8}) is:
\begin{equation}
	\eta^{max}_{SC} = \dfrac{16\sqrt{\dfrac{\tau}{2-\tau}}\cos^{3}N-\tau-4(\tau+2)\cos^{2}N+2}{(1+2\cos(2N))(\tau-2)},\label{eq:9}
\end{equation}
where $N = \dfrac{1}{3}\cos^{-1}\left(-\sqrt{\tau(2-\tau)}\right)$. Now, we obtain expression for the $\Omega$ function by using Eqs. (\ref{eq:31}), (\ref{eq:21}) and Eq. (\ref{eq:9}) in Eq. (\ref{eq:7}) and optimize the resulted $\Omega$ function expression with respect to $z$ that results into
\begin{widetext}
\begin{equation}
 z^{*} = \left[\tau+\dfrac{\tau\left[\tau-2+4(2+\tau)\cos^{2}A-16\left(\sqrt{\dfrac{\tau}{2-\tau}}\cos^{3}A\right)\right]}{2(\tau-2)(1+2\cos(2A))}\right]^{1/3},\label{eq:10}
\end{equation}
where $A = \dfrac{1}{3}\cos^{-1}\left(-\sqrt{1-\eta_{c}^2}\right)$. The expression for the efficiency at maximum $\Omega$ function can be evaluated by using Eq. (\ref{eq:10}) into Eq. (\ref{eq:8}) as:
\begin{equation}
	\eta_{SC}^\Omega = \dfrac{(K-1)(2K^2-(1-\eta_{c})(K+1))}{(1-\eta_{c})-(1+\eta_{c})K^2},\label{eq:11}
\end{equation}
where 

$ K = \left[1-\eta_{c}-\dfrac{(1-\eta_{c})(-1-4(\eta_{c}-3)\cos^{2}A-\eta_{c}-16\sqrt{\dfrac{1-\eta_{c}}{1+\eta_{c}}}\cos^{3}A)}{2(1+2\cos(2A))(1+\eta_{c})}\right]^{1/3}.$

In order to get a clear insight into efficiency at maximum $\Omega$ function, we expand Taylor's series in Eq. (\ref{eq:11}) at near-equilibrium conditions, and will obtain
\begin{equation}
	\eta_{SC}^\Omega = \left(-\dfrac{9}{2}+\dfrac{11\sqrt{3}}{4}\right)\eta_{c}+\dfrac{1}{144}(8339-4804\sqrt{3})\eta_{c}^2+\dfrac{5}{1728}(-179246+103503\sqrt{3})\eta_{c}^3+...\label{eq:12}
\end{equation}
\end{widetext}
To compare the efficiency at maximum $\Omega$ function with that of the maximum work efficiency, we will derive maximum work efficiency by optimizing Eq. (\ref{eq:21}) w.r.t. $z$ and will get $z^*= \tau^{1/3}$. Therefore, the resulting expression of maximum work efficiency is,
\begin{equation}
	\eta_{SC}^{MW} = \dfrac{3(1-\eta_{c})^{1/3}-3+\eta_{c}}{(1-\eta_{c})^{1/3}-1-\eta_{c}}\label{32}.
\end{equation}
In Fig. \ref{fig:2}, for sudden compression stroke, we represent the efficiency at maximum $\Omega$ function (Eq. (\ref{eq:11})) and maximum work efficiency (Eq. (\ref{32})) with solid red and dashed brown curves respectively. We observe that the efficiency attains unit value for the sudden compression stroke. In the inset of this figure, we plot the difference between these two efficiencies, i.e., $\Delta$ = $\eta_{SC}^{\Omega}$ - $\eta_{SC}^{MW}$ with dashed orange curve and notice that the difference between $\eta_{SC}^{\Omega}$ and $\eta_{SC}^{MW}$ gives a positive result, conveying $\eta_{SC}^{\Omega}>\eta_{SC}^{MW}$.

\subsection{Sudden expansion stroke} \label{sb2}

As the mentioned case is opposite to the previous one, we proceed as we did earlier. In this case, we take the expansion stroke (C to D) to be sudden ($\lambda_{CD}$ = $(\omega^2_{c}+\omega^2_{h})/{2 \omega_{c} \omega_{h}}$) and the compression stroke (A to B) as an adiabatic process ($\lambda_{AB}$ = 1). The expressions for absorbed heat and work within a high-temperature regime are, respectively as \cite{singh2024asymmetric}:
\begin{equation}
	Q_{h} = \dfrac{1}{\beta_{h}}\left(1-\dfrac{\tau}{z}\right),\label{34}
\end{equation}
\begin{equation}
	W_{SE}^{HT} = \dfrac{1}{\beta_{2}}(z-1)\left(\dfrac{\tau}{z}-\dfrac{1+z}{2}\right),\label{eq:22}
\end{equation}
We obtain the corresponding efficiency's expression as:
\begin{equation}
	\eta^{HT}_{SE} = \dfrac{(z^{2}-2\tau+z)(z-1)}{2(\tau-z)}.\label{eq:13}
\end{equation}
We derive the upper bound on efficiency by optimizing Eq. (\ref{eq:13}) with respect to $z$, which gives the following cubic equation,
\begin{equation}
	2z^3-3z^2\tau+\tau(2\tau-1)=0. \label{c}
\end{equation}
Again, by using the concept of casus irreducibilies, we obtain the solution of the above equation as 
\begin{equation}
	z^* = \dfrac{\tau}{2} + \tau \cos \left(\frac{1}{3} \cos^{-1}\left[\frac{2-(4-\tau)\tau}{\tau^2}\right]\right). \label{d}
\end{equation}
Now, by using Eq. (\ref{d}) in Eq. (\ref{eq:13}), the expression of maximum achievable efficiency in sudden expansion stroke ($\eta_{SE}^{max}$) is evaluated as follows \cite{singh2024asymmetric}:
\begin{equation}
	\eta^{max}_{SE} = \dfrac{(2-\tau-2F)(4F(\tau+1)-\tau(6-\tau)+4F^2)}{16F-8\tau},\label{eq:14}
\end{equation}
where $F = \tau \cos \left(\frac{1}{3} \cos^{-1}\left[\frac{2-(4-\tau)\tau}{\tau^2}\right]\right)$. Now, by inserting Eqs. (\ref{34}), (\ref{eq:22}), and (\ref{eq:14}) in Eq. (\ref{eq:7}), we derive the expression of $\Omega$ function for sudden expansion stroke and then we optimize the obtained expression with respect to $z$ that gives the following result,

\begin{widetext}
\begin{equation}
	z^{*} = \left[\tau+\dfrac{\tau(\tau-2+2\tau\cos B)(3\tau-6+4(1+\tau)\cos B+2\tau\cos(2B))}{32\cos B-16}\right]^{1/3},\label{eq:15}
\end{equation}
where $B = \dfrac{1}{3}\cos^{-1} \left( \dfrac{2+(\tau-4)\tau}{\tau^{2}} \right)$. We derive the efficiency at the optimal value of $\Omega$ function in sudden expansion case by using Eq. (\ref{eq:15}) into Eq. (\ref{eq:13}) as: 
\begin{equation}
	\eta_{SE}^{\Omega} = \dfrac{(1-C)[2(\eta_{c}-1)+C(C+1)]}{2(C+\eta_{c}-1)},\label{eq:16}
\end{equation}
where	

$	C = \left[1-\eta_{c}+\dfrac{1}{32\cos D-16}(-3-4\cos D(\eta_{c}-2)-2\cos 2D(\eta_{c}-1)-3\eta_{c}) \right. 
	\left. (1-\eta_{c})(-1-\eta_{c}-2\cos D(\eta_{c}-1))\right]^{1/3},$

and $D = \dfrac{1}{3}\cos^{-1} \left( \dfrac{\eta_{c}^{2}+2\eta_{c}-1}{(\eta_{c}-1)^{2}}\right)$.
\end{widetext}

The Taylor's series expansion of the efficiency at maximum $\Omega$ function in sudden expansion stroke (Eq. (\ref{eq:16})) up to the cubic order in terms of Carnot efficiency is represented as

\begin{widetext}
\begin{equation}
	\eta_{SE}^{\Omega} = \left(-\dfrac{9}{2}+\dfrac{11\sqrt{3}}{4}\right)\eta_{c}+\dfrac{1}{36}(1414-815\sqrt{3})\eta_{c}^2+\dfrac{1}{432}(-93262+53853\sqrt{3})\eta_{c}^3+...\label{eq:17}
\end{equation}
\end{widetext}

With the help of Eq. (\ref{eq:22}), we establish the expression for the efficiency at maximum work output for the sudden expansion stroke as
\begin{equation}
	\eta_{SE}^{MW} = \dfrac{3\left[1-(1-\eta_{c})^{2/3}\right]-2\eta_{c}}{2\left[1-(1-\eta_{c})^{2/3}\right]}\label{33}.
\end{equation}
To analyze the performance of the engine under sudden expansion stroke, in Fig. \ref{fig:2}, we plot the efficiency at maximum $\Omega$ function (Eq. (\ref{eq:16})) and the efficiency at maximum work output (Eq. (\ref{33})) with solid green and dashed violet colors, respectively. We observe that the efficiency in the sudden expansion case achieves a value of only one-half. The reason for such efficiency behavior is as follows: during a sudden expansion stroke, the frequency associated with the harmonic oscillator changes suddenly and the system does not remain in its instantaneous state. Non-adiabatic transitions occur among the energy levels, and the system moves to a non-equilibrium state. The state deviates from the diagonal form due to the formation of off-diagonal elements in the density matrix, leading to coherence \cite{liu2023application,streltsov2017colloquium,shi2020quantum}. The additional energy cost of creating these coherences gets stored in the working material. This extra energy, stored as parasitic energy, is transferred to the heat reservoirs and the phenomenon is known as the quantum friction \cite{kosloff2017quantum,plastina2014irreversible,ccakmak2017irreversible,latune2021roles,alecce2015quantum,ccakmak2016irreversibility}.

Additionally,  in the inset of Fig. \ref{fig:2}, we show the difference between $\eta_{SE}^{\Omega}$ and $\eta_{SE}^{MW}$ ($\Delta^{'}$) with solid blue curve. We notice that the difference $\Delta^{'}$ always gives a positive result, again indicating that the efficiency at maximum $\Omega$ function is always greater than the efficiency at maximum work output, i.e., $\eta_{SE}^{\Omega}$ $ > $ $\eta_{SE}^{MW}$.

Further, we compare the efficiency at maximum $\Omega$ function of asymmetric quantum Otto engine (QOE) with that of symmetric one having both work processes as either adiabatic or sudden-switch. From Ref. \cite{PhysRevE.106.024137}, the efficiency at maximum $\Omega$ function for adiabatic case in high-temperature regime is
\begin{equation}
	\eta_{Adi}^{\Omega} = 1-\sqrt{\dfrac{(2-\eta_{c})(1-\eta_{c})}{2}}. \label{m}
\end{equation}
The corresponding efficiency for sudden-switch case is  
\begin{equation}
	\eta_{SS}^{\Omega} = \dfrac{(2-H-2\eta_{c}^2)(2-H+2\eta_{c})}{2(2-H-2\eta_{c})(1+\eta_{c})^2}, \label{n}
\end{equation}
where $ H = \{2(1-\eta_{c})[2+3\eta_{c}^2+2\eta_{c}\sqrt{2(1-\eta_{c})}+\eta_{c}]\}^{1/3} $. In Fig. \ref{fig:2}, solid gray and black curves correspond to efficiency at maximum $\Omega$ function for adiabatic ($\eta_{Adi}^{\Omega}$) and sudden-switch cases ($\eta_{SS}^{\Omega}$), respectively. We observe that the efficiency of the maximum $\Omega$ function is highest in the adiabatic case and lowest in the sudden-switch case. The asymmetric QOE efficiencies for sudden compression and sudden expansion strokes fall between the adiabatic and sudden-switch cases of symmetric QOE. Therefore, we conclude that $\eta_{Adi}^{\Omega}>\eta_{SC}^{\Omega}>\eta_{SE}^{\Omega}>\eta_{SS}^{\Omega}$.

\section{Fractional Loss of Work} \label{sc}

To better analyze the engine's performance, we define a parameter R, which is the fractional loss of work and is given by the ratio of lost work by the engine and extracted work from the engine \cite{singh2020optimal}
\begin{equation}
	R = \dfrac{W_{lost}}{W_{ext}} = \dfrac{\eta_{c}}{\eta}-1, \label{q}
\end{equation}
To derive the expression for fractional loss of work at maximum work output, we use Eq. (\ref{32}) for sudden compression and Eq. (\ref{33}) for sudden expansion in Eq. (\ref{q}) and obtain
\begin{equation}
	R_{SC}^{MW} = \dfrac{(1 -\eta_{c})^{1/3}(3-\eta_{c})+ \eta_{c}(2+\eta_{c})-3}{3-\eta_{c}-3 (1 - \eta_{c})^{1/3}}, \label{w}
\end{equation}
\begin{equation}
	R_{SE}^{MW} = \dfrac{(1 - \eta_{c})^{2/3}(2\eta_{c}-3)-4\eta_{c}+3}{3 (1 - \eta_{c})^{2/3} + 
		2 \eta_{c}-3 }. \label{e}
\end{equation}
\begin{figure}[t]
	\centering
	\includegraphics[width=8.6cm]{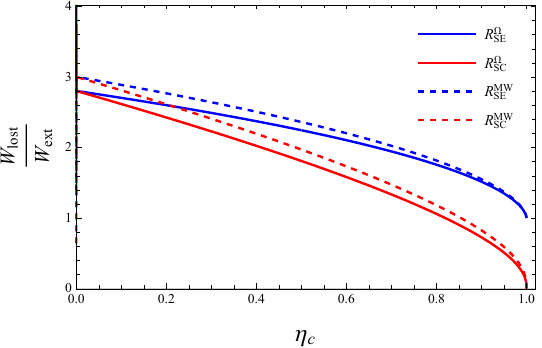}
	\caption{Fractional loss of work as a function of Carnot efficiency $\eta_{c}$. Solid red and blue curves correspond to the fractional work loss at maximum $\Omega$ function during sudden compression and sudden expansion stroke, respectively, while dashed red and blue curves represent the corresponding loss at maximum work (Eqs. (\ref{w}) and (\ref{e})).}\label{fig.4}
\end{figure}
Now, to obtain the expression of fractional work loss at the maximum $\Omega$ function for sudden compression stroke, we use Eq. (\ref{eq:11}), and for sudden expansion case, we use Eq. (\ref{eq:16}) in Eq. (\ref{q}). The expressions for fractional loss of work at the maximum $\Omega$ function for both cases are complicated, therefore, we only plot the obtained expressions. Fig. \ref{fig.4} represents the fractional loss of work as a function of Carnot efficiency. Solid (dashed) red and blue curves show fractional loss of work for sudden compression and sudden expansion at the maximum $\Omega$ function (work), respectively. We analyze that the fractional loss of work is greater for the sudden expansion stroke than for the sudden compression case. That's why the engine's efficiency for sudden compression stroke is greater than that of sudden expansion stroke.

\section{Asymmetric quantum Otto refrigerator} \label{sd}

A refrigerator operates in reverse mode to that of an engine. For the refrigerator, $Q_{c}>$ 0, $Q_{h}<$ 0 and the invested work, $W_{in} = -(Q_{c} + Q_{h})>$ 0. 
Coefficient of performance for a refrigerator is given by
\begin{equation}
	\zeta = \dfrac{Q_{c}}{W_{in}}\label{eq:25}.
\end{equation} 
The $\Omega$ function for refrigerator is the best agreement between the maximum cooling load and minimum lost load and the expression will be as,
\begin{equation}
	\Omega = 2Q_{c}-\zeta_{max}W_{in} \label{1a},
\end{equation}
where $\zeta_{c}$ is Carnot COP and $\zeta_{max}$ ($\leq$ $\zeta_{c})$ is the maximum value of COP that a refrigerator can achieve.

\subsection{Sudden compression stroke}

Proceeding similar to the engine, we take $\lambda_{AB}$ = $(\omega^2_{c}+\omega^2_{h})/2 \omega_{c} \omega_{h}$ and $\lambda_{CD}$ = 1 for this case. The cooling power and the work expressions by using Eqs. (\ref{eq:5}) and (\ref{eq:6}) will be, 
\begin{equation}
	Q_c^{SC} = \dfrac{1}{\beta_{h}}(\tau-z), \quad W_{in}^{SC} = \dfrac{1}{\beta_{h}}(1-z)\left(1-\dfrac{\tau(1+z)}{2 z^2}\right). \label{0a}
\end{equation}
By inserting Eq. (\ref{0a}) in Eq. (\ref{eq:25}), we obtain the following COP expression at high temperature limit,
\begin{equation}
	\zeta_{SC}^{HT} = \dfrac{2z^2(z-\tau)}{(1-z)(2z^2-\tau(1+z))}. \label{Eq.26}
\end{equation}
To obtain the maximum attainable value of COP, we optimize Eq. (\ref{Eq.26}) w.r.t. $z$ and apply the concept of casus irreducibilis in conjunction with the concept of a branch cut by taking $ \tau = \zeta_{c}/(1+\zeta_{c}) $ and obtain,
\begin{equation}
	z_{SC}^{*(max)} = 
	2 \sqrt{\dfrac{\zeta_{c}}{2+\zeta_{c}}} \cos\left[\dfrac{1}{3}\cos^{-1}\left(-\dfrac{\sqrt{\zeta_{c}(2+\zeta_{c})}}{1+\zeta_{c}}\right)+\dfrac{4 \pi}{3} \right] \label{1b}.
\end{equation}
We obtain the maximum COP value by using Eq. (\ref{1b}) in Eq. (\ref{Eq.26}) as,
\begin{widetext}
	\begin{equation}
		\zeta_{SC}^{max} = L = \dfrac{8 G^2 \zeta_{c}\left(\zeta_{c}+2G(1+\zeta_{c})\sqrt{\zeta_{c}/(2+\zeta_{c})} \right)}{(2+\zeta_{c})(1+2G\sqrt{\zeta_{c}/(2+\zeta_{c})})\left[\zeta_{c}(1-2G\sqrt{\zeta_{c}/(2+\zeta_{c})})-8G^2\zeta_{c}(1+\zeta_{c})/(2+\zeta_{c})\right]}, \label{1c}
	\end{equation}
\end{widetext}
where $ G = \sin\left[\dfrac{\pi}{6}-\dfrac{1}{3}\cos^{-1}\left(-\dfrac{\sqrt{\zeta_{c}(2+\zeta_{c})}}{1+\zeta_{c}}\right)\right]. $

With the help of Eqs. (\ref{0a}), (\ref{1c}) and (\ref{1a}), we obtain the expression of $\Omega$ function and optimize the resulted expression w.r.t. $z$ that gives
\begin{equation}
	z_{SC}^{*(\Omega)} = M = \left[\dfrac{L \zeta_{c}}{(1+\zeta_{c})(2+L)} \right]^{1/3}. \label{1d}
\end{equation}
By using above value in Eq. (\ref{Eq.26}), COP at maximum $\Omega$ function for sudden compression stroke is expressed as
\begin{equation}
	\zeta_{SC}^{\Omega} = \dfrac{2 M^2 \left[\zeta_{c} - M(1+\zeta_{c})\right]}{(1-M)\left[\zeta_{c}(M+1)-2M^2(1+\zeta_{c})\right]}. \label{1e}
\end{equation}
The above expression is independent of the system's parameter and depends only on the temperature of the reservoirs (or $\zeta_{c}$). Fig. \ref{fig.6} shows the variation of COP at maximum $\Omega$ function with $\zeta_{c}$ for different cases. In Fig. \ref{fig.6}, we represent COP at the maximum $\Omega$ function (Eq. (\ref{1e})) for sudden compression stroke with solid black curve that shows $\zeta_{SC}^{\Omega}$ increases with $\zeta_{c}$.

\subsection{Sudden expansion case}

Here, we use $\lambda_{AB}$ = 1 and $\lambda_{CD}$ = $(\omega^2_{c}+\omega^2_{h})/2 \omega_{c} \omega_{h}$ and obtain the expression for cooling power and work at high temperature as,
\begin{equation}
\small	Q_c^{SE} = \dfrac{1}{\beta_{h}}\left(\tau-\dfrac{1+z^2}{2} \right), \quad	W_{in}^{SE} = \dfrac{1}{\beta_{h}}(1-z)\left(\dfrac{z+1}{2}-\dfrac{\tau}{z}\right). \label{00a}
\end{equation}
By making use of the Eqs. (\ref{00a}) and (\ref{eq:25}), we get COP at maximum $\Omega$ function for sudden expansion case in high temperature regime as follows,
\begin{equation}
	\zeta_{SE}^{HT} = \dfrac{z\left[2\tau-(z^{2}+1)\right]}{(z-1)\left[z(1+z)-2\tau\right]}. \label{2a}
\end{equation}
Now, for obtaining the maximum COP value, we optimize Eq. (\ref{2a}) w.r.t. $z$ and get the following $z$ value with the help of casus irreducibilies and branch cut concept,
\begin{equation}
	z_{SE}^{*(max)} = \dfrac{\zeta_{c}}{2(1+\zeta_{c})}+\dfrac{\zeta_{c}}{1+\zeta_{c}}\cos\left[\dfrac{1}{3}\cos^{-1}\left(\dfrac{2-\zeta_{c}^2}{\zeta_{c}^2}\right)+\dfrac{4\pi}{3}\right]. \label{2b}
\end{equation}
By inserting Eq. (\ref{2b}) in Eq. (\ref{2a}), we find out the analytical expression for the maximum COP as, 
\begin{equation}
	\zeta_{SE}^{max} = X = \dfrac{J\left[2\zeta_{c}-(1+\zeta_{c})(1+J^2)\right]}{(J-1)\left[J(J+1)(1+\zeta_{c})-2\zeta_{c}\right]}, \label{2c}
\end{equation}
where
$ J= \dfrac{\zeta_{c}}{2(1+\zeta_{c})}-\dfrac{\zeta_{c}\sin\left[\dfrac{\pi}{6}-\dfrac{1}{3}\cos^{-1}\left(\dfrac{2-\zeta_{c}^2}{\zeta_{c}^2}\right)\right]}{1+\zeta_{c}}. $
We obtain the expression for $\Omega$ function by using Eqs. (\ref{00a}) and (\ref{2c}) in Eq. (\ref{1a}), and optimize the obtained expression w.r.t. $z$ that gives
\begin{equation}
	z_{SE}^{*(\Omega)} = Y = \left[\dfrac{X \zeta_{c}}{(1+\zeta_{c})(2+X)} \right]^{1/3}. \label{2d}
\end{equation}
We derive the expression of COP at maximum $\Omega$ function by using Eq. (\ref{2d}) in Eq. (\ref{2a}), 
\begin{equation}
	\zeta_{SE}^{\Omega} = \dfrac{Y\left[2\zeta_{c}-(Y^{2}+1)(1+\zeta_{c})\right]}{(Y-1)\left[Y(1+Y)(1+\zeta_{c})-2\zeta_{c}\right]}. \label{2e}
\end{equation}
\begin{figure}
	\includegraphics[width=8.6cm]{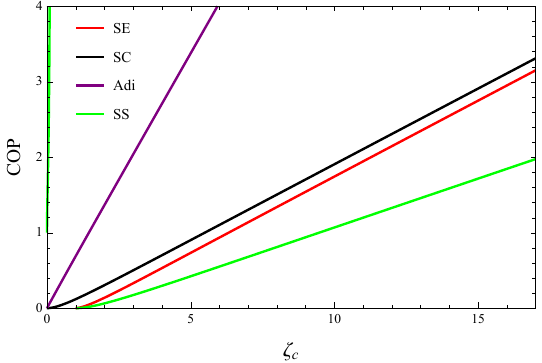}
	\caption{Variation of coefficient of performance at maximum Omega function with Carnot COP $\zeta_{c}$. Black and red curves correspond to COP at maximum $\Omega$ function for sudden compression (Eq. (\ref{1e})) and sudden expansion (Eq. (\ref{2e})), respectively, while purple and green curves show COP at maximum $\Omega$ function for adiabatic and sudden-switch cases.} \label{fig.6}
\end{figure}
We plot COP at the maximum $\Omega$ function for sudden expansion stroke in Fig. \ref{fig.6} with solid red curve. For the comparison purpose, we use the symmetric refrigerator results of COP at maximum $\Omega$ function for adiabatic and sudden-switch cases from Ref. \cite{PhysRevE.106.024137} which are expressed as 
\begin{equation}
	\zeta_{Adi}^{\Omega} = \dfrac{\zeta_{c}}{\sqrt{(2+\zeta_{c})(1+\zeta_{c})}-\zeta_{c}}, \label{1aa}
\end{equation}
and
\begin{equation}
	\zeta_{SS}^{\Omega} =\dfrac{ P\left(1+P(1+\zeta_{c})-\zeta_{c}\right)}{(1-P)\left(P(1+\zeta_{c})-\zeta_{c}\right)}, \label{1ab}
\end{equation}
where $ P =  \left[\dfrac{\zeta_{c}(2\sqrt{2\zeta_{c}(1+\zeta_{c})}-3\zeta_{c}-1)}{(1+\zeta_{c})(2\sqrt{2\zeta_{c}(1+\zeta_{c})}-3(1+\zeta_{c}))}\right]^{1/2} $.
We plot Eqs. (\ref{1aa}) and (\ref{1ab}) in Fig. \ref{fig.6} with solid violet and green curves, respectively. From Fig. \ref{fig.6}, we observe that COP at maximum $\Omega$ function is greater for sudden compression stroke as compared to the sudden expansion stroke and similar to the asymmetric Otto engine, for refrigerator, we have: $\zeta_{Adi}^{\Omega}>\zeta_{SC}^{\Omega}>\zeta_{SE}^{\Omega}>\zeta_{SS}^{\Omega}$.

\section{Conclusion} \label{se}

In this article, we examined the optimal performance of an asymmetric quantum harmonic Otto engine and refrigerator under the $\Omega$ function. First, we derived the analytical expression for efficiency and COP at the maximum $\Omega$ function for the sudden compression stroke. We then repeated our analysis to assess the optimal efficiency and COP during the sudden expansion stroke. Our findings revealed that the efficiency at the maximum $\Omega$ function for sudden compression stroke tends to unity while giving only one-half value for sudden expansion stroke. Furthermore, we compared the obtained optimal $\Omega$ function efficiency with maximum work efficiency and concluded that the former always surpasses the latter. Next, we evaluated the performance of the asymmetric Otto engine and refrigerator in relation to their symmetric counterparts while operating at the maximum $\Omega$ function. We found that the optimal efficiency and COP were highest in the adiabatic case and lowest in the sudden-switch scenario. The efficiencies and COPs of the sudden compression and sudden expansion cases fall between these two extremes. Additionally, for asymmetric engines, we calculated the fractional loss of work for both the sudden compression and sudden expansion scenarios. We observed that the Otto engine performs better during the sudden compression stroke than the sudden expansion stroke due to the less fractional work loss in the sudden compression case.

\section*{Acknowledgment}

We are grateful to V. Singh for the insightful discussions. Monika expresses sincere gratitude to the Government of India for providing financial support through an Institute fellowship at Dr. B. R. Ambedkar National Institute of Technology Jalandhar. 

\appendix
\begin{widetext}
\section{Casus irreducibilis}

While solving the cubic equations, there may arise the case of casus irreducibilis when the discriminant $D = 18abcd-4b^{3}d+b^{2}c^{2}-4ac^{3}-27a^{2}d^{2}$ gives the positive result for the cubic equation\cite{barnett2002methods}

\begin{equation}
	ay^{3}+by^{2}+cy+d = 0.
\end{equation}
The above equation can be expressed in the following form

\begin{equation}
	y^{3}+Ay^{2}+By+C = 0,
\end{equation}
where $A = b/a$, $B = c/a$ and $C = d/a$. We express the solution to the above equation through trigonometric functions as:

\begin{equation}
	y = -\dfrac{A}{3}+\dfrac{2}{3}\sqrt{A^{2}-3B}\cos\left[\dfrac{1}{3}\cos^{-1}\left(-\dfrac{2A^{3}-9AB+27C}{2(A^{2}-3B)^{3/2}}\right)\right].
\end{equation}
We will follow the above procedure to solve the cubic equations for the cases which are discussed in the present paper.

\subsection{Sudden Compression Case}
\label{appendixA}

During sudden compression, the discriminant of cubic equation

\begin{equation}
	z^{3}-\dfrac{3\tau z}{(2-\tau)}+\dfrac{2\tau^{2}}{(2-\tau)} = 0,
\end{equation}
will be $D =  108\tau^{3}(2-\tau)(1-\tau)^{2}>$ 0, $A = 0$, $B = -3\tau/(2-\tau)$, $C = 2\tau^{2}/(2-\tau)$.

\subsection{Sudden Expansion Case}
\label{appendixB}

During sudden expansion stroke, discriminant D for the equation 
\begin{equation}
	z^{3}-\dfrac{3\tau z^{2}}{2}+\dfrac{\tau(2\tau-1)}{2} = 0,
\end{equation}
is $D = 108\tau^{2}(2\tau-1)(1-\tau)^{2}>$ 0, $A = -3\tau/2$, $B = 0$, $C = (2\tau-1)\tau/2$.
   
\end{widetext}
   
    \bibliographystyle{apsrev4-2}
    \bibliography{P1.bib}
    
\end{document}